\documentclass[prb,aps,12pt,nofootinbib]{revtex4}
\newcommand{\be}{\begin{eqnarray}}
\newcommand{\ee}{\end{eqnarray}}

\begin{document}

\title{Fundamental Theory of Statistical Particle Dynamics}
\author{Gene F. Mazenko }
\affiliation{The James Franck Institute and the Department of Physics\\
The University of Chicago\\
Chicago, Illinois 60637}
\maketitle


\centerline{Abstract}

We introduce a fundamental theory for the kinetics of systems
of classical particles.
The theory represents a unification of kinetic theory,
Brownian motion
and field theory.  It
is  self-consistent and is the dynamic generalization of the
functional theory of fluids in equilibrium.
This gives
one a powerful tool for investigating the existence of
ergodic-nonergodic transitions near the liquid-glass transition.



\newpage

\section{Introduction}

We present here a theory\footnote{The title of this paper suggests that the theory is more widely applicable than to Smoluchowski dynamics. This is true. The basic development goes through for Newtonian dynamics, Fokker-Planck dynamics and trapped systems. However, presenting the theory in its most general form makes the development more complicated. Instead, I have chosen to present the theory in its simplest application, the case of Smoluchowski dynamics. The case of Newtonian dynamics will be available soon (S. Das and G. Mazenko).} for the dynamics of  classical particles
which solves the chronic problem of self-consistency.
This theory unites the desirable
elements of kinetic theory\cite{KT}, Brownian motion\cite{BM}, and
modern field theory\cite{FT}.

Kinetic Theory is one of our oldest\cite{ESMKT}
theoretical disciplines.  Despite
its many successes it has never  been constructed in a
fully self-consistent form\cite{self-consistency}. Thus one of the most famous approximations
in all of science, the
{\it stosszahlansatz} of Boltzmann\cite{Stosszahl} and the treatment
of the collision
integral in the Boltzmann equation\cite{BEsum} have not been
investigated systematically.
The theory introduced here provides the tools to remedy this situation.

The stosszahlansatz,  also referred to as the assumption of
molecular chaos, is representative
of decoupling approximations appearing
in many\cite{DCA} problems
and characterized as uncontrolled by the approximation police.
Of particular current interest is the validity of mode coupling
theory
(MCT)\cite{MCA} used in theories of the
liquid glass transition\cite{GTS}.  We should be able to answer
the question:  Is the liquid-glass transition accompanied by
an ergodic-nonergodic
(ENE) transition?\cite{ENE}
The
construction\cite{KTMCT} of mode
coupling models using traditional kinetic theory is
ad hoc and short of convincing.
It has been completely ineffective in exploring corrections to
conventional mode coupling theory.
The field has moved away from  kinetic theory treatments and turned
instead to field theoretical models\cite{FTM}
 where one has the promise of perturbative
control.  Thus we recently introduced the random diffusion
model\cite{RDM}   which can support an ergodic-nonergodic transition\cite{ENE}
at one loop order.  However going to two-loop order
one finds that the system can not sustain the ENE solution. The problem with
such models, compared to microscopic models organized in terms of a
pair potential, is that the short-distance structure is not treated
naturally.  In the case of colloidal systems, there has been some
convergence on the Dean-Kawasaki (DK) model\cite{DK} as the simplest field theoretic
model that describes the kinetics  of the colloidal systems operating
under Smoluchowski\cite{SD1}
dynamics.
It has been difficult,
for technical reasons, to establish whether the Dean-Kawasaki
model supports an ENE transition even at one loop
order.

It is demonstrated here how mode coupling theory\cite{KAWREAR}
 naturally occurs
in the application of our theory to colloidal
systems governed by Smoluchowski dynamics.  The approach,
which allows for
compatible approximations for higher order correlation
functions\cite{HOCF}, is
applicable to a large set of dynamical systems,
\cite{RDYT} reversible\cite{DDYT} and dissipative,
including
Newtonian\cite{ND}, Fokker-Planck\cite{FPD}  and
Smoluchowski\cite{SD1} dynamics.
The theory is organized in terms of a coupling to time and space
dependent external fields.  This allows for great flexibility
in using functional methods in developing various  types of
perturbation theory.

A key point is that the equilibrium equal-time fluid structure\cite{ETL}
has been understood from a self-consistent field theoretical point of view
for a long-time.  The work presented here is the natural extension
to the dynamic regime of the beautiful diagrammatic/functional
development for the static properties.
The theory in the static case proceeded first through the
introduction of graphical methods by  Mayer\cite{Mayer} and
others\cite{ETLO}
in the 1940's and then greatly profited from graphical
resummation techniques\cite{GRT} which
were subsequently supplemented by functional methods\cite{FMESM}
as discussed below.  Of particular interest here is the functional
formulation of Percus\cite{PYi,ETL} which cleanly connects with the
widely applied self-consistent approximations named
Percus-Yevick
\cite{PYE} and hypernetted chain\cite{HNC} which fit
prominently into the tool kit of anyone
studying the statics of fluids.  The theory, which has been applied
to a large variety of systems,  is in the form of nonlinear
integral equations connecting the radial (pair) distribution function,
the direct
correlation function and the Ornstein-Zernike relation\cite{OZR}.

Why has it taken 40 years to extend the static theory to the dynamic regime?

Part of the answer is connected with the difficulties in developing
an efficient field theoretical\cite{ECFT} description for classical field
dynamics.  It was not until the work of Martin, Siggia and Rose\cite{MSR},
introducing field doubling via conjugate response fields, and its
generalization by Janssen\cite{Jan}, De Dominicis\cite{DD2},
Graham\cite{Graham} and others\cite{MSRplus},
that we had a self-consistent treatment for field-theoretic models like the
time-dependent Ginzburg Landau models\cite{TDGL} and all of
the models representing dynamic critical phenomena
universality classes\cite{HH}.
In organizing these dynamical theories it is important to
carefully incorporate
causality.
This is related to the issue of the proper treatment\cite{JAJZJ}
of the Jacobian of
the transformation from a Langevin description in terms of noise to the
path integral description in terms of physical
fields. There are some apparent ambiguities in
determining the Jacobian.  This Jacobian  was identified at least qualitatively
in the early work of Onsager and Machlop\cite{OM1}.
The treatment of the Jacobian has led to a fascinating set of extensions
of the theory to include the topics such as
ghost fermions\cite{GF}, supersymmetry\cite{SS},
Onsager's Reciprocity relations\cite{ORR}  and
Jarzynski/fluctuation theorems\cite{JFT}.
We intend to return to these topics in future work since they can
be explored in the case of particle models of the type studied
here.
Kinetic theory
is complicated\cite{ROSET} compared to conventional
field theories because the
collective variables, particle density and phase-space density,
are distributions (sums of $\delta$-functions)
not smooth fields.  This leads to
nonlinear constraints like
\be
\rho (x_{1})\rho (x_{2})= \rho(x_{1})\delta (x_{1}-x_{2})
+ \textrm{two particle terms}.
\ee

One important aspect of the kinetic theory problem is that
static development in terms of the density
is strongly nongaussian\cite{IGF}.
By this we mean that
the density fluctuations of an ideal gas are not gaussian
and one has a set of irreducible vertex
functions which can not be treated as small.
The connected vertex functions for an ideal gas are not small.
We return to this important
point below.

\section{Smoluchowski Dynamics}

Let us begin by defining the dynamical
system of interest.  Consider a system of $N$ particles with configurations
specified by the coordinates $R_{i}$
which satisfy the equations of motion
\be
\dot{R}_{i}= DF_{i}+\eta_{i}
\label{eq:2}
\ee
where the particles experience force
\be
F_{i}=-\frac{\partial}{\partial R_{i}}U(R),
\ee
with total potential
\be
U(R)=\frac{1}{2} \sum_{i\neq j}V(R_{i}-R_{j})
\label{eq:4}
\ee
where we choose $V(0)=V^{\prime}(0)=V^{\prime\prime}(0)=0$ and we have suppressed vector labels to unclutter
the equations.
There is a noise source $\eta_i$ for each coordinate
which is taken to be gaussian  with variance
\be
\langle\eta_{i}(t)\eta_{j}(t')\rangle =2k_{B}T D\delta (t- t ')\delta_{ij}
\ee
where $D$ is a diffusion coefficient.
It is conventional to develop kinetic theory in terms of the phase-space
density $f(1)=\sum_{i=1}^{N}\delta (x_1-R_{i}(t_1))\delta (p_1-P_{i}(t_1))$ and
its cumulants.  For the case of Smoluchowski dynamics this would
suggest using the particle density $\rho (1)=\sum_{i=1}^{N}\delta (x_1-R_{i}(t_1))$
as the Martin-Siggia-Rose (MSR) field.
The key to our development
is that we break with tradition and treat the particle coordinates
as our MSR fields with  accompanying conjugate response fields.

In the approach developed here we keep track of degrees of freedom
by coupling to them with external source fields. In principle we can keep
track of all the degrees of freedom in the system.  In practice, however,
we are interested in following a small set of collective variables
we label $\Phi$.  For this system, the density $\rho$ is essential
since it governs the static equilibrium behavior and, from the point
of view developed here, is always included in $\Phi =(\rho ,\ldots)$.
For reasons that will be developed below we must also include
in $\Phi$ a field $B$ which is constructed and interpreted
below.
The set of collective variables treated ($\Phi =(\rho , B, \ldots)$),
is flexible and controlled by pairing
each observable with a conjugate external field ($H =H_{\rho},
H_{B},\ldots $).  The set $\Phi$ must include the fields $\rho$
and $B$ since we need both to control and manipulate the interactions
in the system.  One can include other observables in the problem,
like the potential energy density,  but they play a more
passive role in the development.

We take advantage of the fact that, while
the density is strongly nongaussian, the positions
$R_{i}$, may be more
profitably thought of as gaussian variables. Therefore, in developing our theory we do not work in the Langevin description, but instead work in the MSR representation. (As discussed in Appendix A, one has at least three approaches to choose from: Langevin, Fokker-Planck and Martin-Siggia-Rose.) The generator of cumulants in the MSR representation is given by $W_N[H]$ which is related to the $N$-particle partition function by
\be
e^{W_N[H]}=Z_{N}[H]=\int \prod_{i=1}^{N}\bigg[{\cal D}(R_{i}){\cal D}(\hat{R}_{i})
d^{d}R_{i}^{(0)}\bigg] P_0(R_{i}^{(0)} )
e^{-A_{R}}e^{H\cdot \Phi}
\label{eq:9}
\ee
where we have a probability distribution $P_0(R_{i}^{(0)} )$
governing the system at the initial
time $t_{0}$.  The shorthand notation $H\cdot \Phi$
means
$\int d^{d}x_{1}\int_{t_{0}}^{\infty}dt_{1}
\left[H_{\rho}(x_{1},t_{1})\rho (x_{1},t_{1})
+H_{B}(x_{1},t_{1})B (x_{1},t_{1})+\ldots\right]$.
In most of our discussion here we assume the system
is in equilibrium initially and the initial distribution
for a set of $N$ particles is canonical:
\be
P_0[R_0]=e^{-\beta U (R_{0})}/Z_{0}
\ee
where $U$ is the
potential energy defined by Eq.(\ref{eq:4}) and $\beta$ is the inverse temperature.
The MSR action for the problem is given by
\be
A_{R }=\int_{t_{0}}^{\infty}dt_{1}
\sum_{i=1}^{N}\left[\hat{R}_{i}(t_{1})k_{B}T D \hat{R}_{i}(t_{1})
+i\hat{R}_{i}(t_{1})\cdot\left (\dot{R}_{i}(t_{1})
-D F_{i}(t_{1})\right) \right]
+A_{J}
\label{eq:8}
\ee
where the
contribution to the action $A_{J}$ is from the notorious
Jacobian\cite{JAJZJ}. (The steps leading from the Langevin description to the MSR field theory description are discussed in Appendix \ref{app:Langevin}.)
The Jacobian plays a crucial role in this kinetic  problem and
is defined by
\be
J=\textrm{det} \frac{\delta \eta_{i}(t')}{\delta R_{j}(t)}
\nonumber
\ee
\be
=\textrm{det} \frac{\delta }{\delta R_{j}(t)}\left(
\frac{\partial R_{i}(t')}{\partial t'}
-D F_{i}(t')\right)
\nonumber
\ee
\be
=\textrm{det} \delta_{ij}\left[\frac{\partial }{\partial t'}\delta (t-t')
-D\frac{\partial F_{i}(t')}{\partial R_{i}(t')}\delta (t-t')
\right].
\ee
Exponentiating to write the Jacobian as a contribution
to the action gives\cite{JZJTHET}
\be
A_{J}=
-\ln J
=-\int_{t_{0}}^{\infty} dt ~ \theta (0)\left(\sum_{i=1}^{N}
D\frac{\partial F_{i}(t)}{\partial R_{i}(t)}\right )
\label{eq:13}
\ee
where $\theta (0)=1/2$. Together, Eqs.(\ref{eq:9}),
(\ref{eq:8}) and (\ref{eq:13}) define the problem of interest.

Now we want to make a separation of the degrees of freedom into two
groups;  one group consists of some collective variables to be chosen,
and the second group consists of all the rest of the degrees of freedom.
The first step in this separation is to rewrite Eq.(\ref{eq:13})
in terms of the particle density,
$\rho (1)=\sum_{i=1}^{N}\delta (x_1-R_{i}(t_{1}))$.
We find
\be
A_{J}=
\theta (0)\int _{t_{0}}^{\infty}dt\int d^{d}xd^{d}y
D \nabla_{x}^{2}\rho (x,t)V(x-y)\rho (y,t)
+ \textrm{constant}
\nonumber
\ee
\be
=\int_{t_{0}}^{\infty} dt\int d^{d}xd^{d}y B_{J}(x,t)V(x-y)\rho (y,t)
+ \textrm{constant}
\ee
where we have defined the quantity
\be
B_{J}(x,t) = \theta (0)D\nabla_{x}^{2}\rho (x,t)
\ee
and the constant can be absorbed into the normalization
of the partition function.

Next, notice
that the dynamic part of the interaction contribution to the
action can be rewritten in the form
\be
D\int_{t_{0}}^{\infty}dt\sum_{i=1}^{N}i\hat{R}_{i}(t)
F_{i}(t)=\int_{t_{0}}^{\infty}dt\int d^{d}xd^{d}y B_{0}(x,t)
V(x-y)\rho (y,t),
\ee
where
\be
B_{0}(x,t)=D\sum_{i=1}^{N}i\hat{R}_{i}(t)\cdot\nabla_{R_{i}}
\delta ( x-R_{i}(t)).
\ee
We can then combine this contribution to the action with the contribution
from the Jacobian to obtain the dynamic part of the interaction
in the form
\be
A_{I}=\int d1d2 B(1)V(12)\rho (2)
\ee
where $B(1)=B_{0}(1)+B_{J}(1)$ is the field discussed above,
$\int d1 =\int_{t_{0}}^{\infty}dt_{1}d^{d}x_{1}$,
and
\be
V(12)=V(x_{1}-x_{2})\delta (t_{1}-t_{2}).
\ee
Writing things out explicitly, the conjugate field is given by
\be
B(1) =D\sum_{i=1}^{N}\left[(\hat{R}_{i}i\nabla_{1}
+\theta (0)\nabla_{1}^{2})\right]
\delta (x_{1}-R_{i}(t_{1})).
\ee
We can then write the partition function given by Eq.(\ref{eq:9})
in the form
\be
Z_{N}=\int \prod_{i=1}^{N}\bigg[{\cal D}(R_{i}){\cal D}(\hat{R}_{i})
d^{d}R_{i}^{(0)}\bigg]P_0(R_{i}^{(0)})
e^{-A_{0}-A_{I}+H\cdot\Phi}
\ee
where $A_{0}$ is the quadratic part of the action
excluding the quadratic contribution to the
initial probability distribution
\be
A_{0 }=\int_{t_{0}}^{\infty}dt_{1}
\sum_{i=1}^{N}\left[\hat{R}_{i}k_{B}TD  \hat{R}_{i}
+i\hat{R}_{i}\cdot\dot{R}_{i} \right].
\label{eq:19}
\ee
Notice that we have constructed things such that the coordinates
are constrained to have the values $R_{i}^{(0)}$ at
$t=t_{0}$.  We then average over these values.  Here we are
explicitly treating the case where the system is in equilibrium
at $t=t_{0}$, but more general situations are clearly compatible
with the development.
The
interaction part of the action (including the initial probability distribution)
is given in the compact form
\be
A_{I}=\frac{1}{2}\sum_{\alpha,\nu}\int d1d2
\Phi _{\alpha}(1)\sigma _{\alpha\nu} (12)\Phi _{\nu}(2)
\ee
where the Greek labels range over  $\rho$ and $B$ and we introduce
the interaction matrix
\be
\sigma_{\alpha ,\nu} (12)=(-\beta V(12))
\left[ \hat{\rho}_{\alpha}\hat{\rho}_{\nu}\delta(t_1-t_0)
-\beta^{-1}\left(\hat{\rho}_{\alpha}\hat{B}_{\nu}
+\hat{B}_{\alpha}\hat{\rho}_{\nu}
\right)\right]
\label{eq:IM}
\ee
where we have introduced the useful notation
\be
\hat{\rho}_{\alpha}=\delta_{\alpha , \rho}
\ee
and
\be
\hat{B}_{\alpha}=\delta_{\alpha , B}.
\ee
The canonical partition function can be written in the
convenient form
\be
Z_{N}=Tr^{(N)} e^{-A_{I}+H\cdot\Phi}
\label{eq:23}
\ee
where we have introduced the average
\be
Tr^{(N)}{\cal O} =
\int \prod_{i=1}^{N}\bigg[{\cal D}(R_{i}){\cal D}(\hat{R}_{i})
d^{d}R_{i}^{(0)}\bigg] P_0(R_{i}^{(0)}) e^{-A_{0}}{\cal O}(R).
\ee
Notice that the single-particle contribution to the
action, $A_{0}$, (Eq.(\ref{eq:19})), is included in the weight in
$Tr$.
Thus the class of problems of interest are defined in terms of
a path-integral formulation.

Note that the case where there is a strong external potential
acting on the system of particles is easily treated within
the development.  Suppose that the total force acting in
Eq.(\ref{eq:2}) on particle $i$ is of the form
\be
F_{i}^{T}=F_{i}+F_{i}^{E}
\ee
and the external force is generated by a potential
\be
F_{i}^{E}=-\nabla_{R_{i}}\int d^{d}x U_{E}(x,t)\rho (x,t)
=-\nabla_{R_{i}}U_{E}(R_{i}(t),t).
\ee
An important practical example is the case of optical tweezers
where this external potential or trap  can be taken  to be of the form
\be
U_{E}(x,t)=\frac{1}{2}\kappa (t)\left((x-R_{0}(t))^{2}
\right)
\ee
where $\kappa (t)$ is a controllable amplitude for the
potential and $R_{0}(t)$ is the position of the trap.
If one follows the development of the previous section,
one finds that the external force generates a term in
the action
\be
A_{U_{E}}=\int d1 B(1)U_{E}(1)
\ee
which can be included in the one-body term in the action.
In this case, the initial conditions can be influenced in several ways.
If the trap is turned on for times $t> t_{0}$ there is no
change in initial conditions, while one could prepare the
system in a static trap where one would need to add a term
to the initial potential energy,
\be
U \rightarrow U +\int d^{d}x U_{E}(x,t_{0})\rho (x),
\ee
and $U_{E}(x,t)=U_{E}(x,t_{0})$ for $t<t_{0}$.
There are many other possibilities.  The net result of introducing
this external potential $U_{E}$ that couples to the density
is to physically produce the dynamic coupling to the
field $B(1)$,
\be
H_{B}(1)=U_{E}(1).
\ee

If one is interested in fluctuations in equilibrium
in the presence of a time-independent inhomogeneous
potential $u(x_{1})$, then one makes the replacements $H_{\rho}(1)
=u(x_{1})\delta (t_{1}-t_{0})$ and $H_{B}(1)=u(x_{1})$.
$H_{\rho}$ adjusts the initial condition and $H_{B}$ has the effect
of changing the equation of motion to include the force
due to $u(x)$.

We have succeeded in writing our non-equilibrium problem as
a path-integral characterized by a field-dependent
partition function written in the very compact symmetrical form in the grand canonical ensemble,
\be
Z_{T}[H]=\sum_{N=0}^{\infty}\frac{\rho_{0}^{N}}{N!}
Tr^{(N)} e^{\int d1 H(1)\cdot\Phi (1)}
e^{\frac{1}{2}\int d1 d2 \Phi(1)\cdot\sigma\cdot \Phi(2)},
\label{eq:31a}
\ee
which emphasizes the role of the collective fields $\rho$
and $B$.  We have yet to show that this can be expressed
in a form which produces a self-consistent form of perturbation
theory.

\section{Self-Consistent Development for the Generating Functional}

We now want to rewrite the partition function in a form that allows
us to formally carry out the average in Eq.(\ref{eq:31a}).
We can  use the functional
identity\cite{FI25}
\be
e^{-A_{I}+H\cdot\Phi}=
e^{\hat{A}_{T}}e^{H\cdot\Phi}
\label{eq:25}
\ee
where we define the operators
\be
\hat{A}_{T}=\frac{1}{2}\int d1 d2\sum_{\alpha \beta}
\sigma_{\alpha \beta} (12) \hat{H}_{\alpha}(1)\hat{H}_{\beta}(2)
\ee
and
\be
\hat{H}_{\alpha}(1)=\frac{\delta}{\delta H_{\alpha}(1)}.
\ee
The interaction matrix $\sigma_{\alpha \beta}$
 is given by Eq.(\ref{eq:IM}).
Then, using Eq.(\ref{eq:25}) in Eq.(\ref{eq:23}) gives
\be
Z_{N}
=e^{\hat{A}_{T}}Tr^{(N)} e^{H\cdot\Phi}.
\label{eq:28}
\ee
Next, we restrict the set of fields $\Phi_{i}$ to those that  are one-particle
additive,
\be
\Phi_{i}=\sum_{\alpha =1}^{N}\phi_{i}^{\alpha}
\ee
which is true for the particle density $\rho$ and conjugate field $B$, and
notice that the sum over the degrees of freedom in Eq.(\ref{eq:28})
factorizes into a product of sums over the degrees of freedom of
each particle.  Together, these observations lead to the result
\be
Z_{N}^{(0)}=Tr^{(N)} e^{H\cdot\Phi}=(Z_{1})^{N}
\ee
where the noninteracting partition function for a single
particle is
\be
Z_{1}=Tr^{(1)} e^{H\cdot\phi^{(1)}}.
\ee
Working in the grand canonical
ensemble, the grand  partition function for the interacting problem
is given by
\be
Z_{T}=\sum_{N=0}^{\infty}\rho_{0}^{N}\frac{Z_{N}}{N!}
=\sum_{N=0}^{\infty}\frac{\rho_{0}^{N}}{N!}
e^{\hat{A}_{T}}Z_{N}^{(0)}
\nonumber
\ee
\be
=e^{\hat{A}_{T}}
\sum_{N=0}^{\infty}\frac{\rho_{0}^{N}}{N!}
Z_{1}^{N}
=e^{\hat{A}_{T}}e^{W_{0}}
\label{eq:31}
\ee
where $\rho_{0}$ is the fugacity or bare density and
\be
W_{0}=\rho_{0} Tr^{(1)}~ e^{H\cdot\phi^{(1)}}
=\tilde{T}r~ e^{H\cdot\phi}
\ee
where in the last line we have dropped the particle label on the
trace and the field $\phi$.
The cumulants of the fields $\Phi_{i}$  are generated by
taking functional derivatives of the generating functional
\be
W[H]=ln~ Z_{T}
\ee
with respect to $H_{i}$.
The one-point average in a field is given by
\be
G_{i}=\frac{\delta}{\delta H_{i}}W[H]
\nonumber
\ee
where we have used a compact notation where $i$ labels space, time and
fields $\rho$ or $B$.
Substituting for  $W[H]$ using Eq.(\ref{eq:31}) we find
\be
G_{i}
=\frac{1}{Z}_{T}e^{\hat{A}_{T}} e^{W_{0}}
\frac{\delta}{\delta H_{i}}W_{0}
\nonumber
\ee
\be
=\frac{1}{Z}_{T}\tilde{T}r \phi_{i}e^{\hat{A}_{T}}
e^{\phi\cdot H} e^{W_{0}}
\nonumber
\ee
\be
=\frac{1}{Z}_{T}\tilde{T}r \phi_{i}e^{\hat{A}_{T}} e^{\phi\cdot H}
e^{-\hat{A}_{T}}e^{\hat{A}_{T}}e^{W_{0}}
\label{eq:33}.
\ee
It is not difficult to prove the functional
identity:
\be
e^{\hat{A}_{T}} e^{H\cdot\phi}e^{-\hat{A}_{T}}
=e^{H\cdot\phi}e^{[E+\sum_i F_{i}\hat{H}_{i}]}
\label{eq:34}
\ee
where
\be
E=\frac{1}{2}\sum_{ij}\sigma_{ij}\phi_{i}\phi_{j}
\label{eq:51}
\ee
is a self-interaction contribution
and
\be
F_{i}=\sum_{j}\sigma_{ij}\phi_{j}
\label{eq:45}
\ee
will play an important role as we go along.
Using Eq.(\ref{eq:34}) back in Eq.(\ref{eq:33}), we obtain
\be
G_{i}
=\frac{1}{Z}_{T}\tilde{T}r \phi_{i}
e^{H\cdot \phi}e^{[E+F\cdot\hat{H}]}e^{W[H]}
\nonumber
\ee
\be
=\tilde{T}r \phi_{i} e^{H\cdot\phi +E+\Delta W[H]}
\label{eq:43}
\ee
where
\be
\Delta W[H] =W[H+F]-W[H].
\ee
An interesting check on the theory is to show that the self-interaction $E$ given by
Eq.(\ref{eq:51}) vanishes.  This follows if the potential
is constructed to be
zero at the origin ($V(0)=0$).  Our most important result is given by:
\be
G_{i}=\tilde{T}r \phi_{i} e^{H\cdot\phi +\Delta W [H]}.
\label{eq:45a}
\ee

Another result useful in this description follows from taking the
derivative
of Eq.(\ref{eq:31}) with respect to $\rho_{0}$, following steps similar
to those leading to Eq.(\ref{eq:45a}), and integrating with respect
to $\rho_{0}$ which leads to the result
\be
W[H,\rho_{0} ]=\int_{0}^{\rho_{0}}dx Tr
e^{H\cdot\phi + W[H+F,x]-W[H,x]}
\label{eq:46}.
\ee
It takes a little calculus to show that the derivative of Eq.(\ref{eq:46})
with respect to $H$ leads to Eq.(\ref{eq:45a}).
Eq.(\ref{eq:46}) can be rewritten in terms of the more
fundamental identity
\be
\frac{\partial}{\partial \rho_{0}}Z_{T}[H,\rho_{0}]
=\tilde{T}r e^{H\cdot\phi}Z_{T}[H+F,\rho_{0}].
\ee

What about response functions?  The response of the density
to an external potential $U_{E}$ which couples to the density is
given by
\be
\chi_{\rho\rho}(12)=\frac{\delta }{\delta U_{E}(2)}\langle \rho (1)\rangle
=\frac{\delta }{\delta H_{B}(2)}\langle \rho (1)\rangle
\nonumber
\ee
\be
=G_{\rho B}(12)=\frac{\delta }{\delta H_{\rho}(1)}\frac{\delta }{\delta H_{B}(2)}
W[H].
\ee

We must now work to show why Eq.(\ref{eq:45a}) is very desirable.
There are several ways one can use Eq.(\ref{eq:45a}) to build an
approximate theory.  In comparison with the static theory
one would guess that density expansions would be the most
successful.  This may be so, but working with expansions
in the pair interaction are conceptionally simpler and
more direct. It seems clear that in developing density
expansions one will be able to make contact with the hypernetted
chain and  Percus-Yevick approximations. There appears much one
can do about coupling constant renormalization.
We return to discuss density expansions
elsewhere.

The
dependence of the theory  on the pair potential is controlled by the quantity
$\Delta W[H]=W[H+F]-W[H]$.  We can expose the dependence on the potential by
constructing the functional Taylor-series expansion
\be
\Delta W[H] =\sum_{i}F_{i}\frac{\delta}{\delta H_{i}}W[H]
+\sum_{ij}\frac{1}{2}F_{i}F_{j}\frac{\delta^{2}}{\delta H_{i}\delta H_{j}}W[H]+\cdots
\ee
and we can conveniently introduce the set of cumulants:
\be
G_{ij\ldots k}=\frac{\delta}{\delta H_{i}}\frac{\delta}{\delta H_{j}}
\ldots \frac{\delta}{\delta H_{k}}
W[H]
\ee
to obtain
\be
\Delta W[H] =\sum_{i}F_{i}G_{i}
+\sum_{ij}\frac{1}{2}F_{i}F_{j}G_{ij}
+\sum_{ijk}\frac{1}{3!}F_{i}F_{j}F_{k}G_{ijk}
+\ldots
\ee
with $F_{i}$ given by Eq.(\ref{eq:45}).
Clearly, in this form we can take $\Delta W$ to be a functional of $G_{i}$.
One can then use functional differentiation to express higher
order cumulants in terms of the one- and two-point
correlation functions $G_{i}$ and $G_{ij}$.  One has, for example, the
manipulation expressing the three-point cumulant in terms
of lower order objects,
\be
G_{ijk}=\frac{\delta}{\delta H_{k}}G_{ij}
=\sum_{mnp} -G_{im}G_{jn}G_{kp}
\Gamma_{mnp},
\ee
where the irreducible three-point vertex is given as a functional
derivative of the two-point irreducible vertex
\be
\Gamma_{ijk}=\frac{\delta}{\delta G_{k}}\Gamma_{ij}
\ee
and $\Gamma_{ij}$ is precisely the matrix inverse of
the two-point cumulant
\be
\sum_{k}\Gamma_{ik}
G_{kj}=\delta_{ij}.
\label{eq:65}
\ee
The beauty of the modern field theoretical development
is that an approximation for the two-point vertex as a functional
of the $G_{i}$ and $G_{ij}$ generates self-consistent approximations
for all higher order correlation functions.  The method is set up
to carry out various types of renormalization like replacing the bare
interactions with effective interactions.  This will be exploited
elsewhere.  We expect the situation here to be similar to quantum
many-body theory.  Self-consistency and conservation laws can be
brought together to suggest ways of generating approximations
as done by Kadanoff and Baym with their $\Phi$-derivable\cite{KBPhi}
approximations. A key constraint is equilibrium is the fluctuation-dissipation theorem.

Some of the structure of the theory can be appreciated via the
establishment of
a dynamic generalization of the static Ornstein-Zernike
relation\cite{OZR}.  Starting with the functional equation for the two-point
cumulant, one can use the chain-rule for functional differentiation
to obtain:
\be
G_{ij}=\frac{\delta}{\delta H_{j}}G_{i}
\nonumber
\ee
\be
=\tilde{T}r \phi_{i}\phi_{j}e^{H\cdot\phi +\Delta W}
+\sum_{k}\tilde{T}r \phi_{i}e^{H\cdot\phi}\left(\frac{\delta}{\delta G_{k}}e^{\Delta W}\right)
\frac{\delta}{\delta H_{j}}G_{k}
\nonumber
\ee
\be
={\cal G}_{ij}+\sum_{k}c_{ik}G_{kj}
\label{eq:47}
\ee
where
\be
{\cal G}_{ij}=\tilde{T}r \phi_{i}\phi_{j}e^{H\cdot\phi +\Delta W}
\label{eq:54}
\ee
is a single-particle quantity and we have the memory function\cite{TDDC},
self-energy,  or dynamic direct correlation function given by
\be
c_{ij}=\tilde{T}r \phi_{i}e^{H\cdot\phi +\Delta W}
\frac{\delta}{\delta G_{j}}\Delta W.
\label{eq:48}
\ee
Since $\Delta W$ can be treated as a functional of $G_{i}$ we see
at this stage that we have available a self-consistent theory.
If we define the matrix-inverse
\be
\sum_{k}\gamma_{ik}{\cal G}_{kj}=\delta_{ij}
\label{eq:56}
\ee
then the two-point vertex is given without approximation as
\be
\Gamma_{ij}=\gamma_{ij}+K_{ij}
\label{eq:50}
\ee
where
\be
K_{ij}=-\sum_{k}\gamma_{ik}c_{kj}.
\label{eq:63}
\ee

\section{Noninteracting Smoluchowski System}

The first step in applying this theory is to work out the noninteracting
cumulants for the fields $\Phi =(\rho , B)$.  This calculation for the fundamental objects $R$, $\hat{R}$ is carried out in Appendix \ref{app:Gaussian}. The cumulants for the collective fields are worked out in detail in Appendix \ref{app:Collective}.

The final results are
\be
G^{(0)}_{B\ldots B \rho\ldots\rho}(1,\ldots, \ell, \ell+1, \ldots, n)=
\rho_{0}(2\pi )^{d}\delta \bigg(\sum_{i=1}^{n}q_{i}\bigg)b(1)\ldots b(\ell)
e^{N_{n}}
\label{eq:64}
\ee
where
\be
N_{n}=\frac{1}{2}\bar{D}\sum_{i=1}^{n}\sum_{j=1}^{n}
q_{i}\cdot q_{j}|t_{i}-t_{j}|
\ee
and
\be
b(j)=\bar{D}\sum_{i\neq j=1}^{n}
q_{i}\cdot q_{j}\theta (t_{i}-t_{j})
\ee
and where $\bar{D}=k_{B}T D$. Explicitly, the two-point cumulants are
\be
G^{(0)}_{\rho\rho}(q,q';t,t')=\rho_0(2\pi)^d\delta(q+q')e^{-\bar{D}q^2|t-t'|},
\ee
\be
G^{(0)}_{\rho B}(q,q';t,t')=-\rho_0(2\pi)^d\delta(q+q')Dq^2\theta(t-t')e^{-\bar{D}q^2(t-t')},
\ee
\be
G^{(0)}_{B\rho}(q,q';t,t')=-\rho_0(2\pi)^d\delta(q+q')Dq^2\theta(t'-t)e^{-\bar{D}q^2(t'-t)}
\ee
and
\be
G^{(0)}_{BB}(q,q';t,t')=0.
\ee

Notice that $G_{\rho B}$ is retarded and proportional to
$q^{2}$, while $G_{B\ldots B}=0$.
The results for density cumulants agree with the results
from recent work\cite{VCCA} that shows statistical dynamics of the
density of  noninteracting Brownian
particles can be described by a cubic field theory
where the density is the fundamental field.

\section{Perturbation Theory for the Two-point Cumulant}

The perturbation theory can be organized in terms of the
irreducible vertex functions.
It is clear from the generalized Ornstein-Zernike
 equation, Eq.(\ref{eq:47}),
 that the matrix inverse
of the two-point cumulant is given by
Eq.(\ref{eq:50}) with the matrix $\gamma$ defined by Eq.(\ref{eq:56})
and the self-energy $K_{ij}$ by Eq.(\ref{eq:63}).  To get started, one constructs
the noninteracting two-point cumulant using Eq.(\ref{eq:64}) and one finds the matrix inverses to be given by
\be
\gamma^{(0)}_{B\rho}(12)=-\frac{1}{\rho_{0}Dk_{1}^{2}}
\left(\frac{\partial}{\partial t_{1}}+\bar{D}k_{1}^{2}\right)
\delta (t_{1}-t_{2})
\label{eq:67}
\ee
\be
\gamma^{(0)}_{\rho B}(12)=-\frac{1}{\rho_{0}Dk_{1}^{2}}
\left(-\frac{\partial}{\partial t_{1}}+\bar{D}k_{1}^{2}\right)
\delta (t_{1}-t_{2})
\label{eq:68}
\ee
\be
\gamma^{(0)}_{B B}(12)=-\frac{2}{\rho_{0}Dk_{1}^{2}}\delta (t_{1}-t_{2})
\label{eq:69}
\ee
\be
\gamma^{(0)}_{\rho\rho}=0
\label{eq:70}.
\ee

Working to first order in zero external field,
$\Delta W =\sum_{u}F_{u}G_{u}$ and one has contributions
to ${\cal G}_{ij}$ given by
\be
{\cal G}_{ij}=\tilde{T}r \phi_{i}\phi_{j}(1+\Delta W +\ldots )
\nonumber
\ee
\be
=\tilde{T}r \phi_{i}\phi_{j}(1+\sum_{u}F_{u}G_{u} )
\nonumber
\ee
\be
=G_{ij}^{(0)}+\sum_{k,u}G_{ijk}^{(0)}\sigma_{ku}G_{u}.
\nonumber
\ee
It takes some manipulation to carry out the various contributions, as will be discussed in detail elsewhere, but ultimately
\be
{\cal G}_{ij}=\frac{\bar{\rho}}{\rho_{0}}G_{ij}^{(0)}
\nonumber
\ee
where $\bar{\rho}=\langle\rho\rangle$ is the physical average density,
corrected at first order to be $\bar{\rho}=\rho_{0}/(1+\rho_{0}\beta V(q=0))$.
This follows from the perturbation theory analysis of $G_{i}$.

Turning to the dynamic direct correlation function, we have at first
order
\be
\frac{\delta}{\delta G_{j}}\Delta W
=
\frac{\delta}{\delta G_{j}}\sum_{u}F_{u}G_{u}
\nonumber
\ee
\be
=F_{j}=\sum_{k}\sigma_{jk}\phi_{k}.
\ee
Putting this result directly into the defining equation
for the dynamic direct correlation function, Eq.(\ref{eq:48}),
gives
\be
c_{ij}^{(1)}=\tilde{T}r \phi_{i}\sum_{k}\sigma_{jk}\phi_{k}=
\sum_{k}G_{ik}^{(0)}\sigma_{kj}.
\ee
The contribution to the two-point irreducible vertex is given
by the very simple result
\be
K_{ij}^{(1)}=-\sum_{k,\ell}\gamma^{(0)}_{i\ell}
G_{\ell k}^{(0)}\sigma_{kj}
=-\sigma_{ij}.
\ee

Let us look at the first-order theory for the two-point
correlation function.  It satisfies, where it is understood that
$\rho_{0}$ is replaced by $\bar{\rho}$, the matrix kinetic
equation
\be
G_{ij}=G_{ij}^{(0)}+\sum_{k\ell}G_{ik}^{(0)}\sigma_{k\ell}G_{\ell j}
\label{eq:81}.
\ee
This is a matrix equation which holds for times $t_{i},t_{j}\geq t_{0}$
and the two-point cumulant is written more
explicitly as
\be
G_{ij} \rightarrow (2\pi)^{d}\delta (q_{i}+q_{j})
G_{\alpha_{i}\alpha_{j}}(q_{i},t_{i},t_{j})
\ee
where $\alpha_{i}$ takes on the values $\rho$ and $B$ and the translational
invariance of the system is reflected in the multiplying
$\delta$-function.

Traditionally, there have been two recipes or protocols\cite{KFTP} for
evaluating the two-point cumulant. Both  reduce the problem
to effectively a one-time problem.  The kinetic theory protocol (KTP)
is to treat the problem as an initial value problem with the
system in equilibrium at time $t_{0}$ and we determine the single-time
correlation function $G_{\rho\rho}(q,t_{1}-t_{0})$.  This quantity
is available in the current approach by setting $t_{2}=t_{0}$
in Eq.(\ref{eq:47}) and using Laplace transforms.  Traditionally, one
organizes kinetic theory via the time correlation function
\be
C_{AB}( q, t)=\langle B_{-q}e^{i\hat{L}t }A_{q}\rangle
\ee
where $\hat{L}$ is the Liouville operator\cite{Koop} in the case of Newtonian
dynamics.

In the second protocol, called the field theory protocol (FTP), one takes
$t_{0}\rightarrow -\infty $ and builds up the equilibrium
structure from the noise.  One of the technical advantages of this
approach is that one can maintain time translational invariance
over the time line and there is only one time in the problem,
$t_{1}-t_{2}$.
It is natural to work in terms of time Fourier transforms in this case.
This allows one to understand the causal structure in terms of
properties on the complex plane.  Our theory here is similar
to quantum many body theory where one builds up the equilibrium
correlation using thermal Green's functions\cite{KB}.  A difference
is that in the quantum case in equilibrium one must satisfy
the KMS boundary\cite{KMS} conditions.
Both protocols are included in the development here.  It offers the
opportunity of developing approximations that are internally self-consistent
and one would prefer both procedures to produce the same results.

One can work out the full solution to the set of matrix equations
given by Eq.(\ref{eq:81}) with the simple result
\be
G_{\rho\rho}(q,t_{1},t_{2})=
S(q)\tilde{F}(q,|t_{1}-t_{2}|)
\ee
where
\be
\tilde{F}(q,t)=
e^{-\tilde{D}(q)q^{2}t}.
\ee
The static structure factor\cite{MFT} is given by
\be
S(q)=\frac{\bar{\rho}}{(1+\bar{\rho}\beta V(q))}
\label{eq:71}
\ee
and the physical wavenumber dependent diffusion coefficient
is given by\cite{PDIFF}
\be
\tilde{D}(q)=D\beta\bar{\rho}S^{-1}(q).
\ee
Notice that we can, at this level of approximation, introduce
the notion of an effective potential.
Comparing Eq.(\ref{eq:71}) with the
static Ornstein-Zernike\cite{OZR}
relation we can  identify the effective
interaction
\be
V_{EFF}(q)=-\beta^{-1}c_{D}(q)
\ee
where $c_{D}(q)$ is the physical direct correlation function which is assumed
to be known by other means.  We can, for example, assume that
$c_{D}(q)$ is given in the Percus-Yevick approximation for hard
spheres\cite{PYHS}.
 With this
effective interaction one can work out the results of perturbation
theory in $V_{EFF}(q)$.

One can also use the two protocols discussed above to analyze
Eq.(\ref{eq:81}).  In the KTP one sets $t_{2}=t_{0}$ and notices that
only retarded quantities remain in the kinetic equation
which can be solved directly by taking the Laplace transform.
In the FTP  where one takes $t_{0}\rightarrow -\infty$, ones sees,
after taking the limit, all reference to the equilibrium static
structure is gone and one has time translational invariance.
After Fourier transforming over time, the
equations are reduced to a set of algebraic equations which are
simply  inverted to give the same solution in the frequency  regime.
Inverting the Fourier transform leads back to results
found from the complete two-time solution.

To demonstrate the versatility of the method, consider a system
initially $(t=t_{0})$ in equilibrium at temperature $T_{I}$,
but with noise driving the system at $T$ for $t > t_{0}$.
At first order in perturbation theory, the system still satisfies
Eq.(\ref{eq:81}) but with the $t=t_{0}$ contribution in $\sigma$
at temperature
$T$.  The solution of this problem is only slightly more
complicated than the equilibrium case.  One finally has the solution
\be
G_{\rho\rho}(q,t_{1},t_{2})=
S(q)\tilde{F}(q,|t_{1}-t_{2}|)+(S_{0}(q)-S(q))
\tilde{F}(q,t_{1}-t_{0})\tilde{F}(q,t_{2}-t_{0})
\ee
where $S_{0}(q)$ is the static structure factor at temperature $T_{I}$
and $S(q)$ is the static structure factor at temperature $T$.
Now time-translational invariance is broken, but is restored
as the system decays to equilibrium at temperature $T$.

It was claimed earlier that this method could provide
self-consistent approximations for higher-order cumulants.
Within the first-order theory one can generate expressions
for the triplet correlation functions.  We easily find
for the first order theory that
\be
G_{ijk}=-\sum_{vup}G_{iv}G_{ju}G_{kp}\gamma_{vup}^{(0)}
\ee
where these are the first order $G$s on the right hand side and the zeroth order
three-point vertex is given by
\be
\gamma_{vup}^{(0)}=-\sum_{vup}
\gamma_{iv}^{(0)}\gamma_{ju}^{(0)}\gamma_{kp}^{(0)}
G_{vup}^{(0)}
\ee
where all zeroth order quantities can be evaluated using Eq.(\ref{eq:64})
and Eqs.(\ref{eq:67})-(\ref{eq:70}).

At second order in the effective potential we have two contributions
to the two-point vertex.  The first  piece comes from the self-energy
contribution to the dynamic Ornstein-Zernike equation,
Eq.(\ref{eq:48}), where, keeping the second order terms, we have
\be
\frac{\delta }{\delta G_{k}}\Delta W^{(2)}=
\frac{\delta }{\delta G_{k}}\sum_{ij}F_{i}F_{j}G_{ij}
\nonumber
\ee
\be
=-\sum_{ij}F_{i}F_{j}\sum_{\ell , p}
G_{i\ell}\Gamma_{\ell k p}G_{pj}
\ee
and, in the simplest second order approximation, we replace
the $3$-point vertex with the zeroth order result.
After some
simple manipulations that will be described in detail elsewhere,
the second order contributions to the two-point vertex coming from
the self-energy  can be written in the mode coupling form
\be
\Gamma_{ij}^{MC}=-\frac{1}{2}\sum_{k p \ell n}
\gamma^{(0)}_{ik\ell}\delta G_{kp}
\delta G_{\ell n}\gamma^{(0)}_{jpn}
\label{eq:79}
\ee
where the new element is that the high wavenumber convergence of the integrals
comes not from the three point vertices but from the subtraction
in\cite{FRKTSUB}
\be
\delta G_{kp}= G_{kp}- G_{kp}^{(0)}.
\ee
In some field theoretical treatments\cite{ABL} of the DK model
one obtains memory function kernels which do not vanish in the
noninteracting limit.
To complete the second-order model we must also work out
the second-order terms coming from ${\cal G}$.  This
term is linear in the full two-point correlation function
and if there is an ENE transition at this order, it could
show the stretching phenomena found phenomenologically in
Goetze's $F_{12}$ model\cite{F12}.
The complete second order model will be treated in a separate publication.

\section{Conclusions}

We have presented here a reformulation of kinetic theory which
is self-consistent.  It allows one to study problems
which are difficult to treat using other methods.

1.   We outlined a clean derivation of the mode coupling model at second order
in perturbation theory. We will analyze whether this model
supports ENE transitions elsewhere.

2. The method presented here allows for a systematic method for
analyzing corrections to this
second order result including higher order correlation functions.

3.  The method allows one to treat nonequilibrium problems like
temperature quenches as demonstrated above.

4.  Completely unexplored is the fact that the perturbation theory
has been developed in the presence of space and time dependent
external fields.  Thus this method could be useful in problems
of optical pinning and highly inhomogeneous situations.

5.  It seems likely that this method will be useful in treating
meta- and unstable systems.  A first approach within perturbation
theory is to formulate a dynamical van der Waals theory.
Similarly it seems likely that this approach will be useful in
developing a dynamic theory of melting once a few ideas
from density-functional theory  are integrated into the development.

6.  Our focus here, because of its simplicity, has been on
Smoluchowski dynamics, but as will be discussed elsewhere,
the method developed here can be applied both to Newtonian
dynamics and Fokker-Planck dynamics as well as a broader
class of models\cite{GLE}. In the case of Newtonian dynamics,
this approach offers an alternative to the conventional
development in terms of the Liouville operator.  It may be
important that in this approach it is not necessary to use
basis states in perturbation theory labeled by a continuous
momentum index.

7.  It is clear that we can study using these methods the mapping
of Fokker-Planck dynamics onto Smoluchowski dynamics in the large mass limit.
This is of course an elaborated version
of the calculation leading to Einstein's relation \cite{MAZO}
between
the friction coefficient and the diffusion coefficient.
Of particular interest to us is whether in  integrating out
of the momentum degrees of freedom one generates density dependent
diffusion coefficients.  The effects of such a density dependence
been shown to be physically important  within the random diffusion model.\cite{RDM}.

8.  All of this development is compatible with the
bizarre developments initiated with the introduction
of ghost fermions\cite{GF} into the treatment of stochastic dynamics.
We expect the subsequent developments like supersymmetry\cite{SS}
to manifest themselves as in field theories where one can make
a connection with Onsager's reciprocal relations\cite{ORR} and the
fluctuation theorems\cite{JFT}  found in the strongly nonequilibrium
regime.

9.  Independent of the relevance of point 8, the connection
to the fluctuation theorems and Jarzynski equalities
should  be
explored carefully.

10.  These methods allow one to study one-point quantities
in more complicated situations.  In the calculation outlined
above for homogeneous systems we did not mention the self-consistent determination of the equation
of state because the results are rather dull.
This will not be the case for inhomogeneous systems.

11.  It seems clear that a more quantitative method can be established
if one determines the best way of expanding $\Delta W$ in a density
expansion.  This would make close contact to the work of Percus
and the standard approximations of Percus-Yevick and hypernetted chain.

12.  This method allows us to investigate the claim by Das and Mazenko
\cite{DM2009}
that momenta are associated with a mechanism which cuts off any
possible ENE transition in conventional fluids.

13.  Even in the case of Smoluchowski dynamics, the connection between the theory at
the microscopic level and fluctuating nonlinear hydrodynamics
has not been established.  Can the noninteracting field theory of Ref. \onlinecite{VCCA} be
extended to the interacting regime and
connected up to the microscopic theory treated here?

14. The existence of fluctuation-dissipation theorems is very important in equilibrium. This will be explored in depth elsewhere.

\begin{acknowledgments}
This work was supported by the Joint Theory
Institute and the Department of Physics at the University of Chicago.
The author thanks Professors S. Das, K. Freed, S. Rice
and M. Zannetti for comments and also thanks David McCowan and Paul Spyridis for
comments and help with the manuscript.
\end{acknowledgments}

\newpage
\appendix
\section{Langevin, Fokker-Planck and MSR Dynamics}\label{app:Langevin}
\subsection{Langevin Model}
The Langevin model we study is given by
\be
\frac{\partial R_{i}(t)}{\partial t}
= D F_{i}(t) +\eta_{i}(t)
\label{eq:6}
\ee
which is valid for $t> t_{0}$ and has the initial condition
\be
R_{i}(t_{0})=R_{i}^{(0)}
\ee
governed by the probability distribution
$P_{0}[R_{i}^{(0)}]$.
$F_{i}(t)$ is the force acting on $R_{i}(t)$ and is a local function
of $R_{i}(t)$.
We assume that the noise $\eta_{i} (t)$ is gaussian white noise
with variance
\be
\langle\eta_{i} (t)\eta_{j} (t^{\prime})\rangle
=2k_B T D \delta_{ij}
\delta (t-t^{\prime})
\label{eq:3}
\ee
where $D$ is a diffusion coefficient.
The associated probability distribution is given by
\be
P[\eta ]=N_{0}\exp\bigg[-\int_{t_{0}}^{\infty}dt~\frac{\sum_{i}\eta_{i}^{2}(t)}
{4\bar{D}}\bigg]
\ee
where $N_0$ is a normalization factor. The partition function in an external field is given by
\be
Z[h]=\int {\cal D}\eta P[\eta ] \int {\cal D} R^{(0)} P_{0}[R^{(0)}]
S(h\cdot R )
\label{eq:5}
\ee
where the
time-dependent external field couples to the system via
\be
S(h\cdot R )=\exp\left[\int_{t_{0}}^{\infty}dt\sum_{i=1}^N R_i(t)h_{i}(t)\right]
\ee
and where, via the Langevin equation and its initial condition, the field $R_i$ is
a functional of the noise and the initial condition $R_i^{(0)}$.  Our goal
is to determine the generator of cumulants, $W[h]=\ln Z[h]$.

\subsection{Fokker-Planck Dynamics}
It is convenient to analyze this set of dynamical models using the
Fokker-Planck
description.
We follow here the development of
Kim and Mazenko\cite{KM91}.  If we define the field
\be
g_{\phi}(t)=\prod_{i=1}^{N}\delta
\left(\phi_{i}  -R_{i} (t)\right),
\label{eq:7}
\ee
we may then consider the time correlation functions
\be
G_{\phi ,\phi'}(t)=\langle g_{\phi}(t)g_{\phi '}\rangle
\ee
where the average is over the noise and the initial condition.

Taking the time derivative of $G_{\phi ,\phi'}(t)$ using the chain-rule for
differentiation and the Langevin equation, Eq.(\ref{eq:6}), one is left with
\be
\frac{\partial }{\partial t}G_{\phi ,\phi'}(t)
=-\sum_{i}\frac{\delta }{\delta \phi_{i} }
\Big[ F_{i}(\phi )
G_{\phi ,\phi'}(t)
-\sum_{i}
\langle\eta_{i}(t)g_{\phi}(t)g_{\phi '}\rangle \Big].
\label{eq:10}
\ee
It is then not difficult to show, remembering that $\eta$
is gaussian, that
\be
\langle\eta_{i}(t)g_{\phi}(t)g_{\phi '}\rangle
=-\bar{D} \frac{\delta }{\delta \phi _{i}}
G_{\phi ,\phi'}(t)
\label{eq:11}
\ee
where we have used $\bar{D}=k_B T D$, the result
\be
\frac{\delta R_{i} (t)}{\delta\eta_{i}(t)}
=\frac{1}{2}\delta_{ij}
\ee
which follows from Eq.(\ref{eq:6}) and the assumption that the initial
field configuration is independent of the noise for $t\geq0$,
\be
\frac{\delta R_{i} (t_{0})}{\delta\eta_{i}(t)} =0.
\ee
Using Eq.(\ref{eq:11}) back in Eq.(\ref{eq:10}) we can write
\be
\frac{\partial }{\partial t}G_{\phi ,\phi'}(t)
=D_{\phi}G_{\phi ,\phi'}(t),
\label{eq:A13}
\ee
where the Fokker-Planck operator is defined by
\be
D_{\phi}=D \sum_{i}\frac{\delta }{\delta \phi_{i}}
\left[ -F_{i}(\phi )
+k_{B}T
\frac{\delta }{\delta \phi_{i}}
\right].
\label{eq:14}
\ee
The formal solution to Eq.(\ref{eq:A13}) is
\be
G_{\phi ,\phi'}(t-t_{0})=e^{D_{\phi}(t-t_{0})}G_{\phi ,\phi'}(0)
\nonumber
\ee
\be
=e^{D_{\phi}(t-t_{0})}\left[\delta (\phi -\phi')P_{0}(\phi' )\right].
\label{eq:17}
\ee
Integrating over all $\phi'$ gives the equilibrium probability distribution
\be
P(\phi,t)=\int\mathcal{D}(\phi')e^{D_{\phi}(t-t_{0})}[\delta(\phi-\phi')P_0(\phi')]
\nonumber
\ee
\be
=e^{D_{\phi}(t-t_{0})}P_0(\phi).
\ee
It is easy to see that the equilibrium solution is given by
\be
P_{\phi}(0)\equiv W_{\phi}=\frac{e^{-\beta \mathcal{H}_{\phi}}}{Z}
\ee
where
\be
F_{i}=- \frac{\partial {\cal H}}{\partial \phi_{i}}.
\ee
Eq.(\ref{eq:17}) then takes the form
\be
G_{\phi ,\phi'}(t)
=e^{D_{\phi}t}\left[\delta (\phi -\phi')W_{\phi}\right].
\ee

Any two-time correlation can then be written in the form
\be
C_{AB}(t)=\langle A( R (t))B( R (0))\rangle
\nonumber
\ee
\be
=\int {\cal D}\phi {\cal D}\phi ' A(\phi ')B(\phi )G_{\phi ,\phi'}(t)
\nonumber
\ee
\be
=\int {\cal D}\phi B(\phi )e^{D_{\phi}t}A(\phi )W_{\phi}
\label{eq:22}
\ee
where in the final step we assume the system is in equilibrium.

In our development it is useful to introduce the adjoint Fokker-Planck
operator
\be
\tilde{D}_{\phi}=D\sum_{i}
\left[F_{i}
-k_{B}T \frac{\delta}{\delta \phi _{i}}\right]
\frac{\delta }{\delta \phi_{i}}.
\label{eq:35}
\ee
If the equilibrium average is defined as
\be
\langle A\rangle=\int {\cal D}(\phi )W_{\phi}A(\phi ),
\ee
one can show that
\be
C_{AB}(t)=\int {\cal D}(\phi ) B(\phi )e^{D_{\phi}t}(A(\phi )W_{\phi})
=\langle B(\phi )e^{\tilde{D}_{\phi}t}A(\phi )\rangle.
\ee

\subsection{Multiple time correlations}

Consider the multiple time correlation function
\be
G_{\phi_{0},\phi_{1},\phi_{2},\ldots ,\phi_{n}}
(t_{0},t_{1},t_{2},\ldots ,t_{n})=
\langle g_{\phi_{n}}(t_{n})\ldots g_{\phi_{2}}(t_{2})
g_{\phi_{1}}(t_{1})
g_{\phi_{0}}(t_{0})\rangle
\ee
where $g_{\phi}(t)$ is defined by Eq.(\ref{eq:7}) and we assume
$t_{n}\geq t_{n-1}\geq \cdots \geq t_{2} \geq t_{1}\geq t_{0}$.
Using the same approach as used for treating the two-point quantity we have
\be
\frac{\partial }{\partial t_{n}}G_{\phi_{0},\phi_{1},\phi_{2},\ldots ,
\phi_{n}}(t_{0},t_{1},t_{2},\ldots ,t_{n})
=-\sum_{i}\frac{\delta }{\delta \phi_{i} }
\nonumber
\ee
\be
\times
\Big[D F_{i}
G_{\phi_{0},\phi_{1},\phi_{2},\ldots ,\phi_{n}}
(t_{0},t_{1},t_{2},\ldots ,t_{n})
\nonumber
\ee
\be
-\sum_{i} k_{B}T
\langle\eta_{i}(t_{n})
g_{\phi_{n-1}}(t_{n-1})\cdots g_{\phi_{2}}(t_{2})
g_{\phi_{1}}(t_{1})
g_{\phi_{0}}(t_{0})\rangle \Big].
\label{eq:A10}
\ee
Because of causality, $g_{\phi_{i}}(t_{i})$ for $t_{i} < t_{n}$ is
independent of the noise at $t_{n}$:
\be
\langle\eta_{i}(t_{n})
g_{\phi_{n}}(t_{n})\cdots g_{\phi_{2}}(t_{2})
g_{\phi_{1}}(t_{1})
g_{\phi_{0}}(t_{0})\rangle
\nonumber
\ee
\be
=\bar{D}\langle \frac{\delta g_{\phi_{n}}(t_{n})}
{\delta\eta_{\alpha_{n}}(t_{n})}
g_{\phi_{n-1}}(t_{n-1})\cdots g_{\phi_{2}}(t_{2})
g_{\phi_{1}}(t_{1})\rangle
g_{\phi_{0}}(t_{0})\rangle
.
\ee
The treatment of $\frac{\delta g_{\phi_{n}}(t_{n})}
{\delta\eta_{n}(t_{n})}$ is the same as the two-time case
and we obtain the result
\be
\frac{\partial }{\partial t_{n}}G_{\phi_{0},\phi_{1},\phi_{2},\ldots ,
\phi_{n}}(t_{0},t_{1},t_{2},\ldots ,t_{n})=D_{\phi_{n}}
G_{\phi_{0},\phi_{1},\phi_{2},\ldots ,\phi_{n}}(t_{1},t_{2},\ldots ,t_{n})
\ee
where $D_{\phi}$ is the Fokker-Planck operator given by Eq.(\ref{eq:14}).
This has the formal solution
\be
G_{\phi_{0},\phi_{1},\phi_{2},\ldots ,\phi_{n-1},\phi_{n}}
(t_{0},t_{1},t_{2},\ldots t_{n-1},t_{n})
\nonumber
\ee
\be=e^{D_{\phi_{n}}(t_{n}-t_{n-1})}
G_{\phi_{0},\phi_{1},\phi_{2},\ldots ,\phi_{n-1},\phi_{n}}
(t_{0},t_{1},t_{2},\ldots t_{n-1},t_{n-1}).
\ee
However,
\be
G_{\phi_{0},\phi_{1},\phi_{2},\ldots ,\phi_{n-1},\phi_{n}}
(t_{0},t_{1},t_{2},\ldots t_{n-1},t_{n-1})
=\langle g_{\phi_{n}}(t_{n-1})g_{\phi_{n-1}}(t_{n-1})
\cdots g_{\phi_{2}}(t_{2})
g_{\phi_{1}}(t_{1})
g_{\phi_{0}}(t_{0})\rangle
\nonumber
\ee
\be
=\delta( \phi_{n}-\phi_{n-1})
G_{\phi_{0},\phi_{1},\phi_{2},\ldots ,\phi_{n-1}}
(t_{0},t_{1},t_{2},\ldots t_{n-1}).
\ee
Clearly we can work this out recursively:
\be
G_{\phi_{0},\phi_{1},\phi_{2},\ldots ,\phi_{n-1},\phi_{n}}
(t_{0},t_{1},t_{2},\ldots t_{n-1},t_{n})
=e^{D_{\phi_{n}}(t_{n}-t_{n-1})}\delta( \phi_{n}-\phi_{n-1})
G_{\phi_{0},\phi_{1},\phi_{2},\ldots ,\phi_{n-1}}
(t_{0},t_{1},t_{2},\ldots t_{n-1})
\nonumber
\ee
\be
=e^{D_{\phi_{n}}(t_{n}-t_{n-1})}\delta( \phi_{n}-\phi_{n-1})
e^{D_{\phi_{n-1}}(t_{n-1}-t_{n-2})}\delta( \phi_{n-1}-\phi_{n-2})
G_{\phi_{0},\phi_{1},\phi_{2},\ldots ,\phi_{n-2}}
(t_{0},t_{1},t_{2},\ldots t_{n-2})
\nonumber
\ee
\be
=e^{D_{\phi_{n}}(t_{n}-t_{n-1})}e^{D_{\phi_{n-1}}(t_{n-1}-t_{n-2})}
\ldots e^{D_{\phi_{2}}(t_{2}-t_{1})}
 e^{D_{\phi_{1}}(t_{1}-t_{0})}
\nonumber
\ee
\be
\times \delta( \phi_{n}-\phi_{n-1})\delta( \phi_{n-1}-\phi_{n-2})
\ldots \delta( \phi_{1}-\phi_{0})W_{\phi_{0}}.
\ee

If we introduce the notation
\be
U_{\phi_{n};\phi_{n-1}}(t_{n}-t_{n-1})=e^{D_{\phi_{n}}(t_{n}-t_{n-1})}
\delta( \phi_{n}-\phi_{n-1})
\ee
we can write
\be
G_{\phi_{0},\phi_{1},\phi_{2},\ldots ,\phi_{n-1},\phi_{n}}
(t_{0},t_{1},t_{2},\ldots t_{n-1},t_{n})
=U_{\phi_{n};\phi_{n-1}}(t_{n}-t_{n-1})
U_{\phi_{n-1};\phi_{n-2}}(t_{n-1}-t_{n-2})\cdots
\nonumber
\ee
\be
\times
U_{\phi_{2};\phi_{1}}(t_{2}-t_{1})
U_{\phi_{1};\phi_{0}}(t_{1}-t_{0})W_{\phi_{0}}.
\ee
This is the result we need in developing the path integral approach.

\subsection{Path Integral Form}
How is the partition function in a field related to these multiple-time
correlations?  In $Z[h]$, given by Eq.(\ref{eq:5}), we make the special choice for the external field
\be
h_{i}(t)=\sum_{s=0}^{\ell}[i\lambda_{i}^{s}+h_{i}^{s}]\delta (t-t_{s})
\ee
which amounts to dividing up the time interval into a grid.  Next,
multiply by
\be
\prod_{s=0}^{\ell}e^{-\lambda_{i}^{s}\phi_{i}^{s}}
\nonumber
\ee
and integrate over $\lambda_{i}^{s}$. Then, we have
\be
\int~{\cal D}\lambda \prod_{s=0}^{\ell}e^{-i\lambda_{i}^{s}\phi_{i}^{s}}
Z\bigg[\sum_{s=0}^{\ell}[i\lambda_{i}^{s}+h_{i}^{s}]\delta (t-t_{s}) R_{i}(t_{s})\bigg]
\nonumber
\ee
\be
=\int~{\cal D}\eta P[\eta ]\int~{\cal D} R^{(0)}P_{0}[ R^{(0)}]
\int~{\cal D}\lambda \prod_{s=0}^{\ell}e^{-i\lambda_{i}^{s}\phi_{i}^{s}}
e^{i\lambda_{i}^{s} R_{i}(t_{s})}
\nonumber
\ee
\be
\times
e^{\sum_{s,i}h_{i}^{s} R_{i}(t_{s})}
\nonumber
\ee
\be
=\int~{\cal D}\eta P[\eta ]\int~{\cal D} R^{(0)}P_{0}[ R^{(0)}]
\prod_{s=0}^{\ell}g_{\phi_{s}}(t_{s}) e^{\sum_{s,i}h_{i}^{s} R_{i}(t_{s})}
\ee
where again the $g_{\phi}$ are $\delta$-functions which allow us to make
the replacement
\be
e^{\sum_{s,i}h_{i}^{s} R_{i}(t_{s})}=e^{\sum_{s,i}h_{i}^{s}\phi_{i}^{s}}
=S[h\cdot\phi]
\ee
which comes out from the average over noise and initial
conditions.  We have
\be
\int~{\cal D}\lambda \prod_{s=0}^{\ell}e^{-i\lambda_{i}^{s}\phi_{i}^{s}}
Z\bigg[\sum_{s=0}^{\ell}[i\lambda_{i}^{s}+h_{i}^{s}]\delta (t-t_{s}) R_{i}(t_{s})\bigg]
\nonumber
\ee
\be
=S[h\cdot\phi]G_{\phi_{0},\phi_{1},\ldots,\phi_{n}}(t_{0},t_{1},\ldots,t_{n}).
\ee
We finally obtain the result for the partition function we want by doing the
functional integral over $\phi$:
\be
Z[h]=\int~{\cal D}\phi ~
S[h\cdot\phi]G_{\phi_{0},\phi_{1},\ldots,\phi_{n}}(t_{0},t_{1},\ldots,t_{n})
\label{eq:41}.
\ee
The key here is to notice that we have an explicit expression for the multi-time
correlation function $G_{\phi_0,\phi_1,...\phi_n}(t_{0},t_{1},\ldots,t_{n})$.  Inserting this result into
Eq.(\ref{eq:41}), we obtain
\be
Z[h]=\int~{\cal D}\phi ~
S[h\cdot\phi]
U_{\phi_{n};\phi_{n-1}}(t_{n}-t_{n-1})
U_{\phi_{n-1};\phi_{n-2}}(t_{n-1}-t_{n-2})\cdots
\nonumber
\ee
\be
\times
U_{\phi_{2};\phi_{1}}(t_{2}-t_{1})
U_{\phi_{1};\phi_{0}}(t_{1}-t_{0})W_{\phi_{0}}.
\label{eq:39}
\ee
There are no constraints on the
choice of time slices.
Let us take the slices to be uniformly divided,
$t_{s+1}=t_{s}+\Delta$, and we work in
the limit of small $\Delta$ and large $n$.
This defines the continuum limit
for the theory.
As a check on the development, notice that the normalization
as $h\rightarrow 0$
is preserved:
\be
Z[0]=\int~{\cal D}\phi U_{\phi_{n};\phi_{n-1}}(t_{n}-t_{n-1})
U_{\phi_{n-1};\phi_{n-2}}(t_{n-1}-t_{n-2})\cdots
\nonumber
\ee
\be
\times
U_{\phi_{2};\phi_{1}}(t_{2}-t_{1})
U_{\phi_{1};\phi_{0}}(t_{1}-t_{0})W_{\phi_{0}}
=1.
\ee

The result  given by Eq.(\ref{eq:39}) is naturally interpreted
as a path-integral.  Let us focus on the intermediate time
quantities
\be
U_{\phi_{n};\phi_{n-1}}(t_{n}-t_{n-1})=
e^{D_{\phi_{n}}(t_{n}-t_{n-1})}\delta( \phi_{n}-\phi_{n-1}).
\ee
Using the integral representation for the $\delta$-function
we can diagonalize the the Fokker-Planck operator:
\be
U_{\phi_{n};\phi_{n-1}}(t_{n}-t_{n-1})
=e^{D_{\phi_{n}}(t_{n}-t_{n-1})}\int~d\hat{\phi}_{n}
e^{i\hat{\phi}_{n}( \phi_{n}-\phi_{n-1})}.
\ee
Then, we can evaluate
\be
D_{\phi_{n}}e^{i\hat{\phi}_{n} \phi_{n}}
 =\frac{\delta }{\delta \phi_{n}}
\left[ -F_{n}(\phi )
+\bar{D}
\frac{\delta }{\delta \phi_{n}}
\right]
e^{i\hat{\phi}_{n} \phi_{n}}
=A_{n}^{F} e^{i\hat{\phi}_{n} \phi_{n}}
\ee
where
\be
A_{n}^{F}=-i\hat{\phi}_{n}F_{n}(\phi )
-\bar{D} \hat{\phi}_{n}^{2}
-\frac{\delta }{\delta \phi_{n}}F_{n}(\phi )
\ee
and
\be
U_{\phi_{n};\phi_{n-1}}(t_{n}-t_{n-1})
=\int~d\hat{\phi}_{n}
e^{A_{n}^{F}\Delta }
e^{i\hat{\phi}_{n}( \phi_{n}-\phi_{n-1})}
\nonumber
\ee
\be
=\int~d\hat{\phi}_{n} e^{A_{n}\Delta}
\ee
where
\be
A_{n}=-\bar{D} \hat{\phi}_{n}^{2}
+i\hat{\phi}_{n}\left[ ( \phi_{n}-\phi_{n-1})/\Delta
-F_{n}(\phi )-\frac{\delta F_n}{\delta\phi_n}\right].
\label{eq:A50}
\ee
Putting Eq.(\ref{eq:A50}) back in Eq.(\ref{eq:39}), one has
\be
Z[h]=\int~{\cal D}\phi ~{\cal D}\hat{\phi}~{\cal D}\phi_{0}P_{0}(\phi_{0})
e^{\int_{t_{0}}^{\infty}dt A_{R}}
\ee
where $A_{R}$ is the standard MSR action in the presence
of an external field,
\be
A_{R}=-\bar{D} \hat{\phi}^{2}(t)
+i\hat{\phi}(t)\left[ \dot{\phi}(t)
-F(\phi (t) )\right] +h(t)\phi (t)
-\frac{\delta F_{\phi}(t)}{\delta \phi(t)}.
\ee

\newpage
\section{Gaussian Single-Particle Problem \label{app:Gaussian}}

The noninteracting correlations for a system driven by Smoluchowski dynamics is governed by the MSR action
\be
A_0=\int_{t_0}^{\infty} dt \bigg[\hat{R}(t)\bar{D}\hat{R}(t)
+ i \hat{R}(t)\dot{R}(t)-h(t)R(t)-\hat{h}(t)\hat{R}(t)\bigg]
\ee
where $h(t)$ and $\hat{h}(t)$ are the external source fields and $\bar{D}=k_B T D$. We then have the identities that hold in the range $t_0<t<\infty$,
\be
\int\mathcal{D}(R)\mathcal{D}(\hat{R})d^dR_0 P_0[R_0] \frac{\delta}{\delta R(t)}e^{-A_0}=0
\ee
and
\be
\int\mathcal{D}(R)\mathcal{D}(\hat{R})d^dR_0 P_0[R_0] \frac{\delta}{\delta \hat{R}(t)}e^{-A_0}=0.
\ee

Let us begin with the initial condition
\be
P_0[R_0]=\delta(R_0-X_0).
\ee
Evaluating the derivatives of $A_0$, we obtain
\be
2\bar{D}\hat{G}(t)+i\frac{\partial G(t)}{\partial t}=\hat{h}(t).
\ee
and
\be
-i\frac{\partial \hat{G}(t)}{\partial t}=h(t)
\ee
where
\be
G(t)=\langle R(t)\rangle_{X_0}
\ee
and
\be
\hat{G}(t)=\langle\hat{R}(t)\rangle_{X_0}
\ee
where the averages over $R(t)$ and $\hat{R}(t)$ are in the range $t_0<t$. We must now solve these equations to obtain the generating functional.

Using the initial condition $\hat{R}(t_0)=0$, we find that
\be
\hat{G}(t)=-i\int_{t}^{\infty} d\bar{t} h(\bar{t})=\int_{t_0}^{\infty} d\tau g(\tau,t)h(\tau)
\label{eq:B9}
\ee
and
\be
G(t) = X_0 + \int_{t_0}^t d\bar{t} [2i\bar{D}\hat{G}(\bar{t})-i\hat{h}(\bar{t})]
\nonumber
\ee
\be
= X_0 - i\int^{t}_{t_0} d\bar{t} \hat{h}(\bar{t})+ 2\bar{D}\int^{t}_{t_0} d\bar{t} \int_{\bar{t}}^{\infty} d\bar{t'} h(\bar{t'})
\nonumber
\ee
\be
=X_0 +\int_{t_0}^{\infty} d\tau g(t,\tau)\hat{h}(\tau)+\int_{t_0}^{\infty} d\tau C(t,\tau) h(\tau)
\label{eq:B11}
\ee
where
\be
g(t,t')=-i\theta(t-t')
\ee
and
\be
C(t,t')=2\bar{D}\int^t_{t_{0}} d\bar{t} \int^{\infty}_{\bar{t}} d\bar{t'} \delta(t'-\bar{t'}).
\label{eq:B12}
\ee

The generating functional satisfies
\be
\hat{G}(t) = \frac{\delta \ln Z_0(h,\hat{h},X_0)}{\delta \hat{h}(t)}
\label{eq:B14}
\ee
and
\be
G(t) = \frac{\delta \ln Z_0(h,\hat{h}, X_0)}{\delta h(t)}.
\label{eq:B15}
\ee

The generating functional solution to this set of equations -- Eqs. (\ref{eq:B9}), (\ref{eq:B11}), (\ref{eq:B14}) and (\ref{eq:B15}) -- is given by
\be
\ln Z_0(h,\hat{h},X_0) =\frac{1}{2} \int dt \int dt' h(t) C(t,t')h(t')
\nonumber
\ee
\be
+\int dt \int dt' h(t) g(t,t') \hat{h}(t')
+\int dt h(t)X_0.
\ee

The full generator requires averaging over the initial conditions:
\be
Z[h,\hat{h}] = \int d^dR_0 P_0[R_0]e^{\frac{1}{2}h\cdot C\cdot h +h\cdot g\cdot \hat{h}+h\cdot ig\cdot R_0}
\nonumber
\ee
\be
= e^{\frac{1}{2}h\cdot C\cdot h +h\cdot g\cdot \hat{h}}\int d^dR_0 P_0[R_0]e^{h\cdot ig\cdot R_0}.
\ee
All of the equilibrium cumulants can be constructed from Eq.(\ref{eq:B12}) as
\be
C(t,t')=2\bar{D}\int^t_{t_{0}} d\bar{t} \int^{\infty}_{\bar{t}} d\bar{t'} \delta(t'-\bar{t'})
\nonumber
\ee
\be
=2\bar{D}\int^t_{t_{0}} d\bar{t} \theta(t'-\bar{t})
\nonumber
\ee
\be
=2\bar{D}\theta(t-t')\int^{t'}_{t_{0}} d\bar{t}
+2\bar{D}\theta(t'-t)\int^{t}_{t_{0}} d\bar{t}
\nonumber
\ee
\be
=2\bar{D}\theta(t-t')(t'-t_0)+2\bar{D}\theta(t'-t)(t-t_0).
\label{eq:B17}
\ee
\newpage

\section{Collective $\phi$-Correlations}
\label{app:Collective}

We need to evaluate $\phi$-correlations in the non-interacting
case. We start with
\be
W_{0}[H]=\tilde{T}r e^{H\cdot \phi }
\ee
which generates all $\phi =(\rho , B)$ correlations. We proceed by reintroducing the microscopic
sources $h$ and $\hat{h}$ and treating
\be
Z_{0}[H,h,\hat{h}]=\tilde{T}r e^{H\cdot \phi }e^{h\cdot R +\hat{h}\cdot\hat{R}}
\ee
where
\be
h\cdot R = \int_{t_0}^{\infty} dt h(t) R(t).
\ee
Next, we express the $\phi$ in terms of $R(t)$ and $\hat{R}(t)$ as
\be
\phi_{\rho} (1)=e^{-ik_{1}R(t_{1})}
\ee
and
\be
\phi_{B} (1)=-D[(k_{1}\cdot \hat{R}(1)+\theta(0)k_1^2)\phi_{\rho} (1)].
\ee
Let us introduce the operators
\be
\hat{\phi}_{\rho}(1)=
e^{-ik_{1}\frac{\delta}{\delta h(t_{1})}}
\ee
and
\be
\hat{\phi}_{B}(1)=\hat{b}(1)\hat{\phi}_{\rho}(1)
\ee
where
\be
\hat{b}(1)=
-D\bigg[k_{1}\frac{\delta}{\delta \hat{h}(t_{1})}
+k_{1}^{2}\theta(0)\bigg]
\ee
so we may write
\be
e^{H\cdot \phi }e^{h\cdot R +\hat{h}\cdot\hat{R}}=e^{H\cdot \hat{\phi} }e^{h\cdot R +\hat{h}\cdot\hat{R}}
\ee
and
\be
Z_0[H,h,\hat{h}]=e^{H\cdot\hat{\phi}}Z_0[h,\hat{h}]
\ee
where $Z_0[h,\hat{h}]$ was determined in Appendix \ref{app:Gaussian}.

Taking functional derivatives, we can determine all of the noninteracting cumulants of the complete set of densities $\phi$. We have
\be
G_{B\ldots B\rho\ldots\rho}(1,\ldots\ell,\ell+1,\ldots,n)=
\frac{\delta}{\delta H_B(1)}\ldots\frac{\delta}{\delta H_B(\ell)}
\frac{\delta}{\delta H_{\rho}(\ell+1)}\ldots\frac{\delta}{\delta H_{\rho}(n)}
\nonumber
\ee
\be
\times e^{H\cdot\hat{\phi}}e^{\frac{1}{2}h\cdot C\cdot h}e^{h\cdot g\cdot \hat{h}}\int d^d R_0 P_0[R_0] e^{h\cdot ig\cdot R_0}|_{H=h=\hat{h}=0}
\nonumber
\ee
\be
=\hat{\phi}_{B}(1)\ldots\hat{\phi}_{B}(\ell)\hat{\phi}_{\rho}(\ell+1)\ldots\hat{\phi}_{\rho}(n)
e^{\frac{1}{2}h\cdot C\cdot h}e^{h\cdot g\cdot \hat{h}}\int d^d R_0 P_0[R_0] e^{h\cdot ig\cdot R_0}|_{H=h=\hat{h}=0}.
\ee
All $n$-point cumulants have $n$ factors of $\hat{\phi}_{\rho}$ and $\ell$ factors of $\hat{b}$ corresponding to the number of $B$ insertions:
\be
G_{B\ldots B\rho\ldots\rho}(1\ldots\ell,\ell+1,\ldots,n) =\hat{b}(1)\ldots\hat{b}(\ell)\hat{\phi}_{\rho}(1)\ldots\hat{\phi}_{\rho}(n)
e^{\frac{1}{2}h\cdot C\cdot h}e^{h\cdot g\cdot \hat{h}}
\nonumber
\ee
\be
\times \int d^d R_0 P_0[R_0] e^{h\cdot ig\cdot R_0}|_{h=\hat{h}=0}.
\ee
Because the $\hat{\phi}_{\rho}(j)$ are translation operators, it is not difficult to show that
\be
\hat{\phi}_{\rho}(1)\ldots\hat{\phi}_{\rho}(n)F[h(j)]=F[h(j)+L_n(j)]
\ee
where
\be
L_{n}(j)=-i\sum_{s=1}^{n}k_{s}\delta (t_{j}-t_{s}).
\label{eq:C14}
\ee
Thus, we have
\be
G_{B\ldots B\rho\ldots\rho}(1,\ldots\ell,\ell+1,\ldots,n)
\nonumber
\ee
\be
=\hat{b}(1)\ldots\hat{b}(\ell)
e^{\frac{1}{2}(h+L_n)\cdot C\cdot (h+L_n)}e^{(h+L_n)\cdot g\cdot \hat{h}}
\int d^d R_0 P_0[R_0] e^{(h+L_n)\cdot ig\cdot R_0}|_{h=\hat{h}=0}
\nonumber
\ee
\be
=\hat{b}(1)\ldots\hat{b}(\ell)
e^{\frac{1}{2}L_n\cdot C\cdot L_n}e^{L_n\cdot g\cdot \hat{h}}
\int d^d R_0 P_0[R_0] e^{L_n\cdot ig\cdot R_0}|_{\hat{h}=0}
\nonumber
\ee
\be
=\hat{b}(1)\ldots\hat{b}(\ell)
e^{N_n}e^{L_n\cdot g\cdot \hat{h}}
\int d^d R_0 P_0[R_0] e^{L_n\cdot ig\cdot R_0}|_{\hat{h}=0}
\ee
where we have defined
\be
N_{n}=\frac{1}{2}L_{n}\cdot C \cdot L_n.
\label{eq:C16}
\ee

Using the definition of $\hat{b}$, we similarly find that
\be
\hat{b}(j)e^{L_n\cdot g\cdot \hat{h}}=b_n(j)e^{L_n\cdot g\cdot \hat{h}}
\ee
where
\be
b_{n}(j)=-D[k_{j}L_{n}(\bar{t})g(\bar{t}-t_{j})-\theta(0)k_{j}^2]
\nonumber
\ee
\be
=Dk_j\cdot\sum_{s=1}^n k_s \theta(t_s-t_j)-D\theta(0)k_j^2
\nonumber
\ee
\be
=Dk_j\cdot\sum_{s\neq j=1}^n k_s \theta(t_s-t_j).
\ee

In the average over initial conditions, we need
\be
L_n\cdot ig =-i\sum_{s=1}^n k_s
\ee
and
\be
\int d^d R_0 P_0[R_0] e^{L_n\cdot ig\cdot R_0}=
\int d^d R_0 P_0[R_0] e^{-i\bigg(\sum_{s=1}^n k_s\bigg) \cdot R_0}=
(2\pi)^d\delta\bigg(\sum_{s=1}^n k_s\bigg)
\ee
which enforces translational invariance in space. Finally, putting this all together we have
\be
G_{B\ldots B\rho\ldots\rho}(1,\ldots\ell,\ell+1,\ldots,n)
=\rho_0b(1)\ldots b(\ell)e^{N_n}(2\pi)^d\delta\bigg(\sum_{s=1}^n k_s\bigg).
\ee

The argument of the exponential contribution can be put into a more symmetric form. Starting with $N_n$ given by
Eq.(\ref{eq:C16}) and inserting $L_n$ from Eq.(\ref{eq:C14}), we have
\be
N_{n}=-\frac{1}{2}\sum_{i=1}^{n}\sum_{j=1}^{n}k_i k_j C(t_i,t_j).
\ee

The zeroth-order correlation function for the
Brownian coordinates is given by Eq.(\ref{eq:B17}) as
\be
C(t,t')=2\bar{D}\left[\theta (t-t')(t'-t_{0}) +\theta (t'-t)(t-t_{0})\right]
\ee
which, at equal times, reduces to
\be
C(t,t)=2\bar{D}(t-t_0).
\ee
In the zeroth order density correlation functions we find
the quantity
\be
N_{n}=-\frac{1}{2}\sum_{i,j}^{n}k_{i}k_{j}C(t_{i},t_{j})
\ee
with the constraint that $\sum_{i}k_{i}=0$.  This quantity
should be time translationally invariant.

To see this, let us first define
\be
D_{ij}\equiv C_{ii}+C_{jj}-2C_{ij}
\nonumber
\ee
\be
=2\bar{D}\bigg(t_{i}+t_{j}-2[\theta (t_{i}-t_{j})t_{j} +\theta (t_{j}-t_{i})t_{i}]\bigg)
\label{eq:C26}
\ee
\be
=2\bar{D}\bigg(t_{i}[\theta (t_{i}-t_{j})-\theta (t_{j}-t_{i})]
+t_{j}[\theta (t_{j}-t_{i})-\theta (t_{i}-t_{j})]\bigg)
\nonumber
\ee
\be
=2\bar{D}(t_{i}-t_{j}) sgn(t_{i}-t_{j})
\nonumber
\ee
\be
=2\bar{D}|t_{i}-t_{j}|.
\ee
Notice that this result holds if $t\rightarrow t-t_0$ in Eq.(\ref{eq:C26}).

We then have for the argument of the exponential for the $\rho$-$B$ correlation functions
\be
N_{n}=-\frac{1}{2}\sum_{i,j}^{n}k_{i}k_{j}C(t_{i},t_{j})
\nonumber
\ee
\be
=-\frac{1}{2}\left(\sum_{i=1}^{n}k_{i}^{2}C_{ii}
+\sum_{i\neq j}k_{i}k_{j}C_{ij}\right)
\nonumber
\ee
\be
=-\frac{1}{2}\left(\sum_{i=1}^{n}k_{i}^{2}C_{ii}
+\sum_{i\neq j}k_{i}k_{j}\frac{1}{2}
[C_{ii}+C_{jj}-D_{ij}]\right)
\nonumber
\ee
\be
=-\frac{1}{2}\left(\sum_{i=1}^{n}k_{i}^{2}C_{ii}
-\sum_{i=1}^{n}k_{i}^{2}C_{ii}
-\frac{1}{2}\sum_{i\neq j}k_{i}k_{j}D_{ij}\right)
\nonumber
\ee
\be
=\bar{D}\frac{1}{2}\sum_{i\neq j}k_{i}k_{j}|t_{i}-t_{j}|.
\ee
For the special case of $n =2$, we have
\be
N_{2}=-\bar{D}k_{1}^{2}|t_{1}-t_{2}|.
\ee


\begin{thebibliography}{99}
\bibitem{BM}  Brownian motion is interpreted here as a dynamics that
is noise driven.  For an individual Brownian particle located
at position $R$ acted upon by noise $\eta$, one has
$\dot{R}=\eta$. This is clearly the continuum version of the
random walk problem.
\bibitem{self-consistency}
By self consistency we mean that the correlation and response functions of interest satisfy kinetic equations where the collision kernels, self-energies or memory functions can be expressed in terms of the full physical correlation and response functions. In the areas of critical dynamics and the liquid-glass transition, this is essential in generating symmetry breaking solutions. This can generate a multiplicity of solutions not available in bare perturbation theory.
\bibitem{Stosszahl}
 Boltzmann's Stosszahlansatz is a prime example of a decoupling
approximation.  In the collision integral of the Boltzmann
equation he assumed that the two-particle distribution function
factorizes into a product of two one-particle distributions,
$f_{2}\rightarrow f_{1}f_{1}$.
\bibitem{ENE}  The important idea of an ergodic-nonergodic (ENE) transition
exists independent of whether it conforms to the details of mode
coupling theory.  The defining property of an ENE transition is that
as a function of a control parameter there is cross over
from an ergodic phase
($\lim_{t \rightarrow \infty}G_{\rho\rho}(q,t) = 0$)
to a region with non-ergodic kinetics
($\lim_{t \rightarrow \infty}G_{\rho\rho}(q,t) = A^{2}(q)> 0$).
\bibitem{DDYT} Purely dissipative Langevin systems have long
been used to describe the order parameter dynamics in magnetic and
superfluid systems.  There is a balance between the force due
to the gradient of the effective free energy and the persistent noise
in a thermalized system.
\bibitem{IGF} Suppose we have an ideal Newtonian fluid in a field, $u({\bf x})$, that couples to the density.
Next we determine the generating functional with the result
$W[u]=\rho_{0}\int d^{d}x e^{u(x)}$.  If the density is a gaussian
variable, then $W[u]=\int d^{d}x_{1} d^{d}x_{2}
u(x_{1}) G(x_{1}-x_{2})u(x_{2})$.
\bibitem{FI25}
This is just the functional generalization of the standard
matrix identity
\be
\exp\left[\frac{1}{2}\sum_{ij}K_{ij}
\frac{\delta^{2}}{\delta h_{i}\delta h_{j}}\right]
e^{\sum_{i}h_{i}\phi_{i}}=e^{\sum_{i}h_{i}\phi_{i}}
\exp\left[\frac{1}{2}\sum_{ij}K_{ij}\phi_{i}\phi_{j}\right].
\nonumber
\ee
\bibitem{TDDC} In different contexts the kinetic kernels have
different names.  In the field theory protocol the kernels are
typically called self-energies, in the kinetic theory protocol,
where the analysis is in terms of retarded quantities, the kernel is
called a memory function, and in the general case it can be called
a dynamic direct correlation function.
\bibitem{PDIFF}
The decay time in the problem is inversely proportional
to the static structure factor.  This clearly, away from $q=0$,
leads to a slowing down near the first structure factor peak.
\end{thebibliography}
\end{document}